# An Artificial Intelligence Framework for Conflict Mapping and Resolution for Sustainability of Systems

*Apala Chakrabarti*

Independent Researcher

chakrabartiapala@outlook.com

**Abstract:** Early design decisions strongly influence environmental, economic and social outcomes, yet sustainability assessment tools rarely reveal trade-offs among these three pillars. This study presents a framework for Conflict Mapping and Resolution for Sustainability of Systems (CONFARM).

CONFARM consists of four steps: lifecycle documentation, cause-effect mapping, conflict database construction and multi-criteria scoring. A conflict is recorded when a single decision produces positive and negative effects across pillars. Each effect is evaluated using impact magnitude and pillar weight to generate a sustainability ratio. CONFARM may be applied manually or through automated extraction using natural-language processing and large language models.

The method is demonstrated in three sectors representing different data structures and system scales: agriculture (rice and corn), fashion (slow and fast fashion) and energy (nuclear and natural gas). Each system was analysed at increasing conflict densities. Results consistently showed that sustainability scores converged as more conflicts were mapped, indicating stable evaluation across methods. Slow fashion and nuclear systems exhibited relatively higher sustainability performance, while fast fashion and natural gas systems showed lower performance.

CONFARM improves early-stage decision support by making trade-offs explicit and enabling comparative evaluation. It offers a structured approach for cleaner production and scalable sustainability assessment across domains.

## 1. Introduction

According to the Brundtland Report [1], sustainability refers to the capacity to sustain human and ecological systems over time. In design and manufacturing, it is understood as maximising environmental, social, and economic benefits across a product's lifecycle while minimising associated costs.

The sustainability paradigm is conventionally structured around three pillars- environmental, economic, and social [2]. These pillars underpin widely used assessment methods such as Life Cycle Assessment (LCA) and Carbon Footprint Calculators [3-5]. LCA provides quantitative insight into overall environmental performance, while carbon-footprint tools focus primarily on greenhouse-gas emissions. Social Life Cycle Assessment (SLCA) and Life Cycle Costing (LCC) have been developed to extend sustainability evaluation into social and economic dimensions [46-47], and in some cases, these are combined within Life Cycle Sustainability Assessment (LCSA) frameworks [48]. However, even when such integration is attempted, the interactions and conflicts among the pillars are often left implicit rather than explicitly identified and evaluated, particularly during early stages of design when major decisions on long-term sustainability outcomes are determined.

This imbalance limits understanding of cross-pillar interactions and associated trade-offs. Early design and manufacturing decisions determine most lifecycle impacts [6, 7], yet current tools rarely capture the systemic trade-offs among environmental efficiency, cost, and social well-being. Moultrie et al. [6] observed that the lack of an integrated framework constrains evaluation of sustainability in product design, impeding efforts toward cleaner production. Without such integration, conflicting outcomes-such as a process that reduces emissions but increases worker risk or material inequity-often remain undetected until late in development, when corrective actions become costly or infeasible [7].

To address this limitation, the present study proposes a framework for Conflict Mapping and Resolution for Sustainability of Systems (CONFARM),for early-stage detection and integrated evaluation of sustainability-conflicts the environmental, economic, and social pillars. CONFARM employs a structured sequence of cause–effect mapping, conflict database construction, and multi-criteria scoring to identify, classify, and quantify trade-offs across product-lifecycle stages.

A second contribution of this work is the introduction of an AI-assisted implementation that automates conflict extraction and scoring using large-language-model (LLM) techniques (ChatGPT and Gemini). These models analyse lifecycle documentation to identify contradictory effects across sustainability pillars, reducing analysis time and enabling scalable assessment across diverse systems.

The proposed approach is validated through three case studies, one each from the agriculture, fashion, and energy sectors, to demonstrate the adaptability of the proposed framework to systems of varying data richness and complexity. Comparative analysis of manual and AI-assisted implementations demonstrates how linguistic reasoning complements quantitative scoring to reveal sustainability trade-offs that may be overlooked in conventional analyses.

Overall, this research supports cleaner production by bridging qualitative cause-effect reasoning with quantitative evaluation, providing both methodological and computational advancements toward integrated sustainability assessment.

## 2. Background

### 2.1 Sustainability in Product Development

Sustainability has become a central concern in product development since early design decisions determine the majority of a product's long-term environmental, social, and economic impacts. Material selection, process choice, and functional specification established during the concept design stage of product development directly influence energy requirements, waste generation, and social equity throughout the lifecycle of a product [7, 8]. Consequently, decision support at the early stages of design is critical: once a product enters manufacturing, opportunities for sustainable redesign diminish rapidly.

### 2.2 The Three Pillars of Sustainability

The sustainability paradigm is typically described through three interdependent pillars-environmental, economic, and social [9].

- Environmental aspects encompass resource efficiency, energy use, waste generation, and ecosystem impacts.
- Economic aspects involve cost, efficiency, and long-term resilience.
- Social aspects include health and safety, working conditions, community welfare, and equity.

While the three-pillar model is widely cited, its operationalisation in product design remains limited [10]. Environmental and economic indicators are comparatively well developed, whereas social sustainability metrics remain fragmented and qualitative. This imbalance has contributed to the dominance of environmental assessments in industry and research, and results across the three pillars are rarely contrasted or integrated.

### 2.3 Existing Approaches to Sustainability Assessment

Several methods have been introduced to evaluate sustainability across lifecycles. Life Cycle Assessment (LCA) is the most established method for quantitative assessment of environmental impacts from raw-material extraction to end-of-life [11], but it pays limited attention to social and economic factors and requires extensive data, which restricts early-stage use. Carbon-footprint calculators provide simplified estimates of greenhouse-gas emissions [3–5], at the cost of reducing sustainability in terms of a single indicator.

Complementary tools have been developed to address these gaps, such as eco-design toolkits [8], sustainability-maturity models [12], and indices like the Sustainability Compliance Index (SCI) [7]. While these extend analysis beyond purely environmental metrics, they often remain domain-specific or descriptive, with limited ability to quantify cross-pillar interactions and trade-offs.

### 2.4 Limitations in Current Practice

Despite the diversity of tools, sustainability assessment in product design remains pillar-skewed. Economic considerations are usually reduced to cost analysis, and social effects are rarely integrated in a structured way [6, 10]. Furthermore, most assessments measure relative improvements - for example, percentage reductions in emissions or material use, without embedding these results in a holistic, sustainability context. As a result, conflicting effects among pillars often go unrecognised, and decision-makers cannot evaluate whether improvements in one domain undermine another.

### 2.5 Identified Research Gap

The literature consistently highlights the absence of a holistic yet practical framework capable of integrating all three pillars at the early design stages [6, 7]. Environmental assessment methodologies are comparatively mature, while economic aspects are often limited to cost-based evaluation, and social dimensions remain the least analytically treated. As a result, even when all three pillars are considered, they are typically evaluated in isolation rather than in relation to one another. This gap constrains designers from identifying sustainability conflicts that arise from interdependent design and process decisions. Addressing this limitation is essential for advancing cleaner production practices and supporting balanced, evidence-based design evaluation [12].

To guide this study, the following research questions are formulated:
**RQ1.** How can sustainability conflicts that emerge from early-stage design and process decisions be systematically identified across the environmental, economic, and social pillars?
**RQ2.** How can these conflicts be represented and quantified into a sustainability conflict assessment to enable comparison across products, sectors, and levels of data richness?
**RQ3.** To what extent can an AI-assisted implementation of the proposed assessment method support or complement the manual expert-driven process in identifying and evaluating these sustainability conflicts?

In this context, design decisions refer to product-related choices such as material selection, component configuration, and functional specification. Process decisions involve operational choices that affect how the product is made or used, including manufacturing methods, energy sources, workflow settings, logistics, and end-of-life handling. Both decision types can produce sustainability impacts across the three pillars, and these questions drive the development of the proposed framework and structure its validation across the case studies presented in later sections.

## 3. Methodology

### 3.1 Design Research Methodology (DRM)

This study adopts the Design Research Methodology (DRM) [13] to structure theoretical development and empirical validation. DRM provides a systematic basis for design research through iterative refinement between descriptive and prescriptive stages.

Four stages of DRM, corresponding to Type IV research, are applied:

1. Research Clarification (RC): identification of gaps in sustainability-assessment tools, particularly limited integration of social and economic pillars.
2. Descriptive Study I (DS-I): literature review of sustainability standards and multi-pillar assessment approaches to derive framework requirements.
3. Prescriptive Study (PS): development of the conflict-detection framework, integrating environmental, economic and social criteria across lifecycle stages.
4. Descriptive Study II (DS-II): empirical evaluation of CONFARM through case studies, leading to refinement.

This structure ensures conceptual grounding and practical validation remain aligned with design and production realities. In this paper, RC and DS-I form the basis of the literature review and problem framing (Sections 1–2), CONFARM development corresponds to PS, and the case-study application represents DS-II (Sections 3 and 5).

### 3.2 Framework Development

Descriptive Study I showed that existing sustainability assessment approaches, particularly LCA, provide structured procedures for defining system boundaries, organising lifecycle information and characterising impacts [14]. This offers a strong basis for systematic reasoning, but LCA is primarily oriented toward environmental outcomes and does not distinguish when the same design or process decision produces competing effects. The first stage of CONFARM therefore adopts LCA-style lifecycle structuring, while subsequent stages introduce analytical steps to make consequences explicit. These include mapping decision–outcome relationships, identifying where effects diverge and quantifying the resulting tensions. This progression motivates the four-step framework proposed below.

CONFARM was developed as a stepwise process to identify and evaluate sustainability conflicts across the three pillars. It proceeds through four sequential stages as shown in Figure 1:
(1) Product-lifecycle documentation and structuring,
(2) Cause-effect mapping,
(3) Conflict-database construction, and
(4) Multi-criteria scoring and benchmarking.

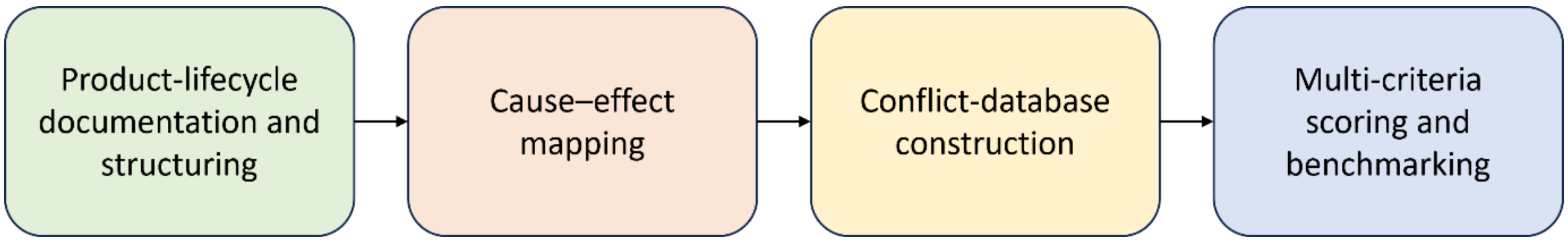

Figure 1: Proposed Framework

### 3.2.1 Product Lifecycle Documentation and Structuring

The first step standardises information across the product lifecycle (PLC) [27-28]. Key data sources include design specifications, process flowcharts, bills of materials, compliance records and environmental reports, retrieved from manufacturer repositories, regulatory portals and sector databases. They record component characteristics, process steps, material and energy inputs, and regulatory requirements, forming the basis for identifying lifecycle decisions and corresponding effects.

Each product is decomposed into five lifecycle stages as seen in Figure 2:

1. Raw–material acquisition,
2. Manufacturing,
3. Transportation and distribution,
4. Use phase, and
5. End–of–life management.

Within each stage, information is organised under three categories:

- Inputs: material and energy flows, resource use, and process conditions.
- Outputs: emissions, waste streams, costs, and social outcomes.
- Decision nodes: design or process interventions with measurable sustainability implications.

Inputs and outputs are documented using Life Cycle Inventory (LCI) practice, where material and energy flows, resource use, emissions and waste are itemised to describe system behaviour [28]. Decision nodes capture interventions with measurable sustainability effects. Design interventions change product attributes such as material or structure, whereas process interventions modify operational choices such as manufacturing method, energy source or waste handling. This enables tracing how interventions propagate through the lifecycle and influence environmental, economic and social outcomes.

When quantitative data are unavailable, qualitative information is converted into categorical scales for consistent evaluation. This structured dataset forms the basis for subsequent cause-effect analysis.

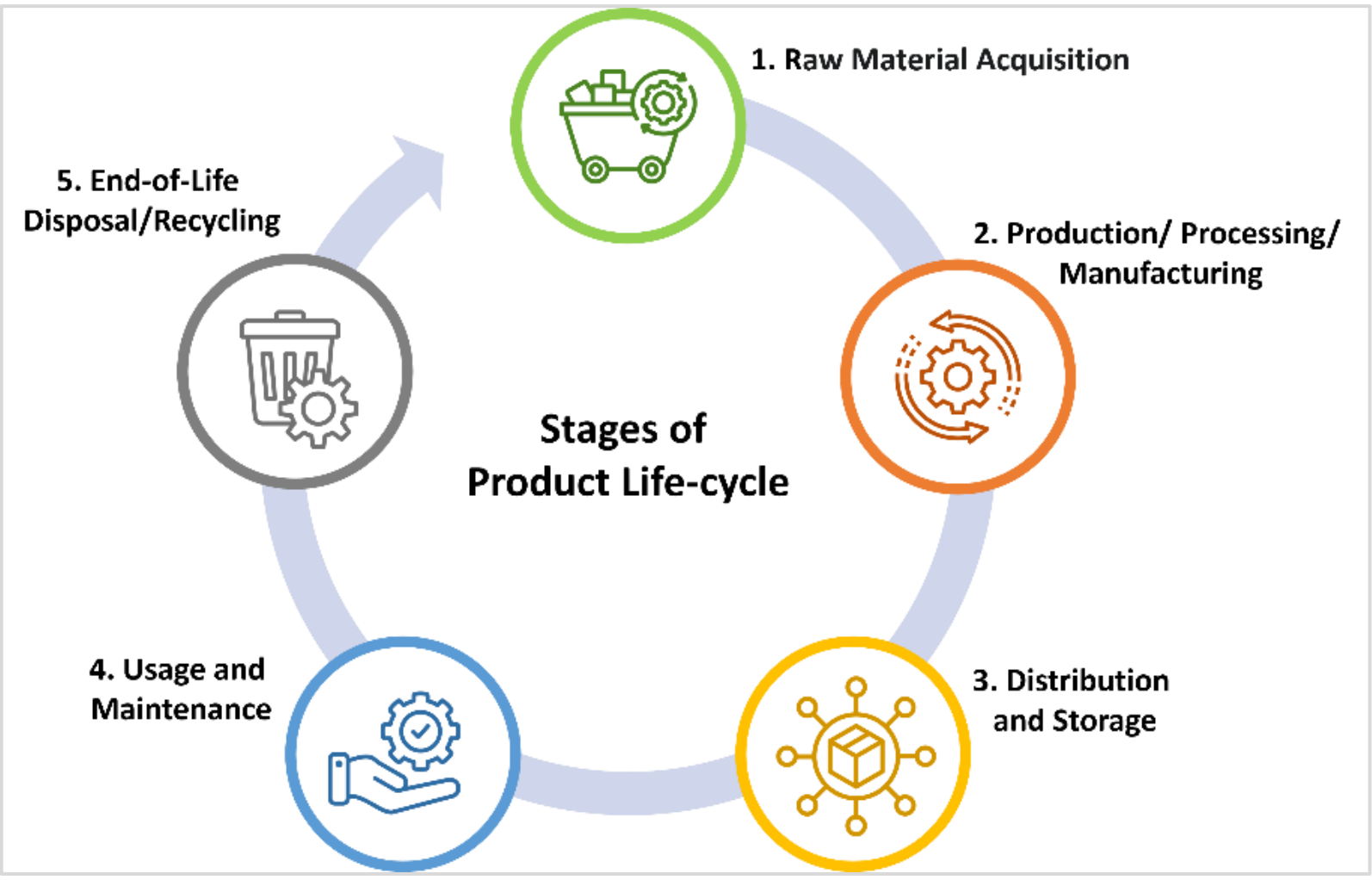

Figure 2: 5 Life-Cycle Stages

### 3.2.2 Cause–Effect Mapping

Cause-effect (CE) mapping traces relationships between decision nodes and sustainability outcomes.

- Cause nodes represent design or process decisions such as material substitution, energy-source selection, or manufacturing-process choice.
- Effect nodes denote resulting sustainability outcomes, classified as positive or negative relative to each pillar's objective.

A conflict is recorded when a single decision produces divergent effects across sustainability pillars. For example, lightweight composites reduce emissions (positive environmental) but increase cost and reduce recyclability (negative environmental/economic). Conflicts are therefore identified whenever one choice creates benefits in one dimension and burdens in another.

Cause–effect mapping traces these relationships through lifecycle documentation. Each decision is linked to the material or process change it introduces, and then to its outcomes. Outcomes are classified as environmental, economic or social, based on effects on resource use and emissions, cost and efficiency, or labour and community conditions. Classification is grounded in the reasoning from decision to consequence rather than generic labels.

The resulting maps reveal interdependencies among lifecycle decisions and form the basis for conflict-database construction. This stage operationalises RQ1 by providing a systematic procedure to detect sustainability conflicts during early design and process decisions.

### 3.2.3 Conflict Database Construction

Conflicts identified through CE mapping are compiled into a structured database that enables traceability, comparison, and quantitative evaluation. Each conflict record includes:

- Lifecycle stage of occurrence,
- Triggering decision or intervention,
- Involved sustainability pillars (environmental, economic, social),
- Associated positive and negative impacts, and
- Type of trade-off (intra-pillar or cross-pillar).

When assessing a new product or process, the decision is documented in the same structure used for existing entries in the database. At this stage, the database is populated through the case studies presented, with each entry recording the decision, the associated material or process change, and the observed environmental, economic and social outcomes.

### 3.2.4 Multi-Criteria Scoring and Benchmarking

The conflict database is evaluated using a multi-criteria scoring method adapted from weighted objectives approaches in decision analysis and sustainability assessment [49]. This is appropriate because sustainability decisions generate different types of effects that cannot be reduced to a single metric. Criteria are taken directly from the documented decision–outcome relationships, where each effect is assigned magnitude and relevance. The method aggregates effects across pillars while retaining traceability of how each outcome contributes to the overall score.

Each effect is assigned two attributes:

- Impact magnitude ($m_i$): severity of the effect, derived from observed outcomes in the case data and classified based on whether the impact is reversible, controllable, or structural.
- Pillar weight ($w_i$): relative importance of each pillar in the given design context, determined through expert judgement during case analysis and consistent with prioritisation practices reported in design for sustainability literature.

The overall sustainability score S is calculated as

$$S = \sum(w_i \times m_i) \qquad [1]$$

This scoring structure is adapted from the regular weighted objectives method [49] widely used in multi-criteria decision analysis, where attributes of different types are normalised and aggregated to support comparative evaluation.

Impact magnitudes are classified based on the scale of intervention required to reverse or mitigate the effect, as shown in Table 1

Table 1: Understanding Impact Magnitudes

| **Magnitude Level** | **How to Identify (Decision Rule)** | **Interpretation** |
|---|---|---|
| **High (0.75)** | The impact is structural or embedded in the product or process and cannot be reversed without redesigning components, reconfiguring workflows, or altering end-of-life pathways. | Requires system-level change. Long-term or widespread consequences if unaddressed. |
| **Medium (0.50)** | The impact can be reduced through targeted adjustments (e.g., material substitution, parameter optimisation, sourcing changes), but may accumulate over time if unmanaged. | Controllable but non-trivial. Mitigation is possible without redesigning the system. |
| **Low (0.25)** | The impact is localised, reversible or limited in scope and can be addressed through operational measures (e.g., process tuning, waste sorting, handling protocols). | Minor or easily reversible. Does not alter system configuration. |

These distinctions are derived from descriptive analysis of the case studies, where the effort and scale of intervention required to address each effect were used to determine magnitude. The numerical values are distributed uniformly over the 0–1 range to provide a consistent comparative basis in the absence of a validated empirical severity scale.

Positive and negative effects are aggregated separately to yield total positive (P) and negative (N) scores. The sustainability ratio R is expressed as

$$R = \frac{N}{(P+N)} \qquad [2]$$

Lower R values indicate higher sustainability performance, while higher values reflect greater unsustainability.

Two derived indicators are proposed for supporting comparative evaluation:

1. Conflict Count: number of distinct conflicts identified through CE mapping-

$$C = |\{c \in \mathbb{C}\}| \quad [3]$$

2. C reflects the breadth of sustainability trade-offs in a system. A high value of C indicates that many points in the lifecycle give rise to competing impacts, even if the individual impacts are individually small. Conflict Magnitude (T) is the total weighted sum of all effects, representing the cumulative strength of sustainability impacts. Each effect has a magnitude $m_i$ (Table 1) and a pillar weight $w_i$. The total magnitude is expressed as:

$$T = \sum_{i=1}^{n} \omega_i m_i \quad [4]$$

$$\mathrm{T} = \mathrm{P} + \mathrm{N} \quad [5]$$

The use of a cumulative sum is justified because the purpose of $T$ is not to distinguish between beneficial and adverse directions, but to measure the overall intensity of sustainability consequences generated by the set of decisions. Even where $\mathrm{P}$ offsets $\mathrm{N}$, their combined magnitude still reflects the system workload in managing sustainability effects (e.g., interventions, monitoring, mitigation, etc.). Therefore, $T$ captures how strong and resource-intensive the trade-offs are, whereas $R$ captures their balance.

Both indicators are required because a system may exhibit:

- many small conflicts (high $C$, low $T$); where individual impacts are minor and dispersed, typically resulting in a low or moderate sustainability ratio $\mathrm{R}$; or
- few but highly consequential conflicts (low $C$, high $T$); where a small number of decisions dominate performance, often producing a high sustainability ratio $\mathrm{R}$ if these impacts are predominantly negative.

These conditions have different implications for design and policy. $C$ indicates where trade-offs are distributed across the lifecycle, $T$ indicates how intense they are, and $\mathrm{R}$ reflects the overall balance of sustainability outcomes. Used together, the three indicators provide a more complete basis for comparison and interpretation.

This stage operationalises RQ1 by providing a systematic procedure to identify sustainability conflicts arising from early-stage design and process decisions across the three pillars. The scores can be interpreted using Table 2.

| **Score Range** | **Sustainability Interpretation** | **Meaning** |
|---|---|---|
| **0.00 – 0.33** | High sustainability | Impacts are negligible or well-managed. No major trade-offs. System design aligns with sustainability goals. |

| Score Range | Sustainability Interpretation | Meaning |
|---|---|---|
| **0.34 – 0.66** | Moderate sustainability | Noticeable trade-offs. Some impacts accumulate across pillars. Improvements are possible through targeted interventions. |
| **0.67 – 1.00** | Low sustainability | Strong or multiple negative impacts. Trade-offs are significant. Prioritise redesign, mitigation, or systemic change. |

The ranges in Table 2 interpret the scores and indicate whether a configuration exhibits high, moderate or low sustainability. These thresholds support benchmarking across cases.

### 3.3 Implementation Pathways

CONFARM can be implemented through two complementary pathways: manual and AI-assisted. Having both implementations addresses RQ3 by enabling comparative evaluation of expert interpretation and automated reasoning in detecting and scoring sustainability conflicts.

## 4. Framework Implementation

The proposed framework can be implemented through two complementary approaches: a manual expert-driven process and an AI-assisted automated process. Both follow the same analytical structure, outlined in Section 3, but differ in how lifecycle data are parsed, mapped and scored.

### 4.1 Manual Method

In the manual approach, all stages- lifecycle structuring, cause-effect mapping, conflict documentation and scoring are all performed through expert review. Product lifecycle documents such as design specifications, process sheets and material records are examined and assigned to the five lifecycle stages. Experts then identify design or process decisions that influence environmental, economic or social outcomes. A conflict is recorded when a single decision produces effects of opposing polarity, and each conflict is scored using the predefined weighting and impact scheme.

Although more time-intensive, the manual method provides key advantages:

- Interpretive accuracy: experts can capture nuanced dependencies and latent social factors that may not be explicitly documented.
- Transparency: every conflict and its rationale are traceable.
- Flexibility: qualitative or incomplete data can be incorporated without loss of consistency.

The manual workflow is particularly effective in early-stage projects or sectors where structured data are limited. Figure 3 depicts the process flow for the manual implementation of CONFARM.

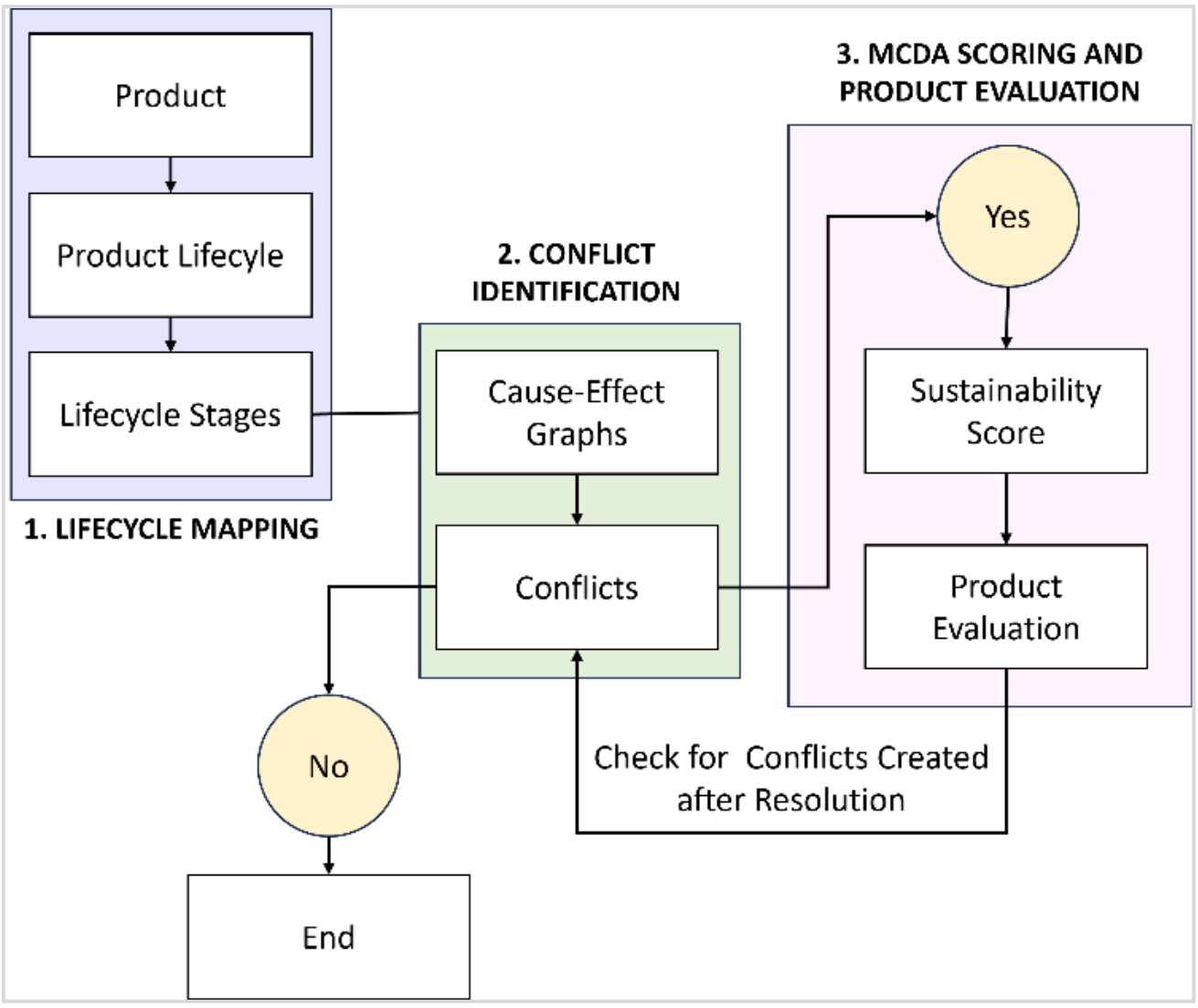

Figure 3: Manual Implementation

The workflow begins by identifying the product and decomposing the lifecycle into stages. Documentation for each stage is reviewed, and cause–effect representations are constructed to link decisions to sustainability outcomes. These representations reveal conflicts when a decision produces both positive and negative effects across pillars. Conflicts are then scored using the weighting scheme to derive a sustainability score. If the score meets the threshold, evaluation continues; otherwise, redesign or alternatives are considered. This sequence ensures consistent application of lifecycle structuring, causal reasoning and conflict evaluation during early-stage assessment.

### 4.2 AI-Assisted Method

The AI-assisted implementation automates framework execution using natural-language processing (NLP) and large-language models (LLMs). PLC documents are uploaded in digital form and parsed automatically to extract relevant design decisions, material data, and process parameters. The algorithm then:

1. Classifies extracted information into the five lifecycle stages.
2. Identifies potential cause–effect relationships linking design choices to sustainability outcomes.
3. Detects conflicts when polarity scores diverge across pillars.

Each conflict is quantified using the same scoring function and weighting scheme as in the manual framework, preserving methodological equivalence. The AI-assisted implementation is illustrated in Figure 4.

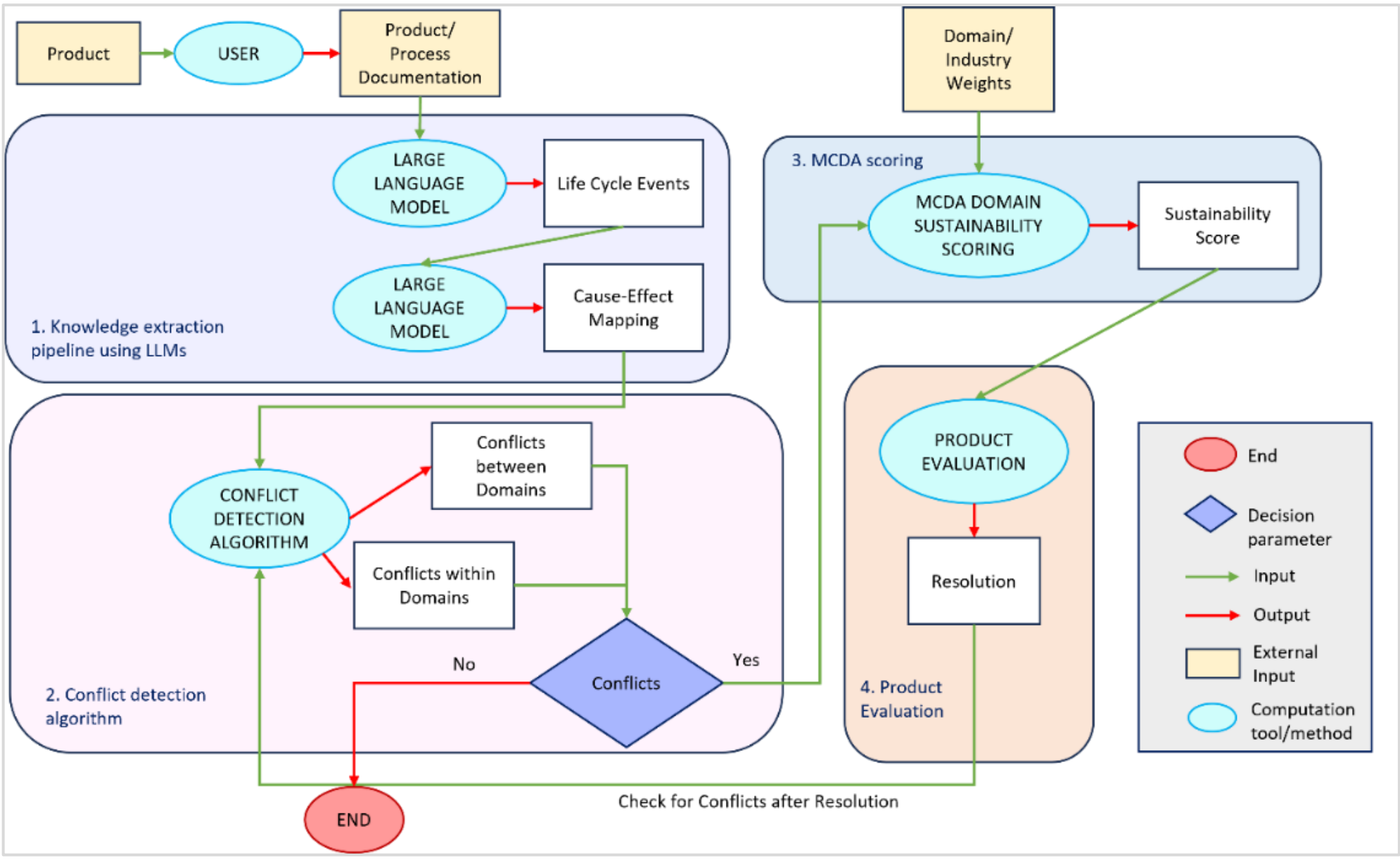

Figure 4: AI Implementation

Lifecycle documents are parsed automatically, and extracted decisions and parameters are classified into the five stages. The algorithm identifies cause-effect links, detects conflicts based on polarity differences, and evaluates each conflict using the established scoring scheme. Scores below the threshold indicate redesign requirements. This route achieves the same analytical results as the manual method but with higher speed and scalability.

The AI-assisted pathway offers several benefits:

- Scalability: large or evolving document sets can be analysed rapidly.
- Consistency: identical logic and scoring are applied across datasets.
- Speed: analysis time is reduced from hours or days to minutes.

However, interpretive precision in social or context-dependent effects still benefits from expert validation, making hybrid use of both methods preferable in complex studies.

### 4.3 Operational Comparison

Table 3 summarises the comparative features of the manual and AI-assisted routes. While both share the same analytical backbone, they differ in efficiency, data requirements, and interpretive capability.

**Table 3.** Comparative features of the manual and AI-assisted implementation routes within the proposed sustainability-conflict-detection framework.

| **Attribute** | **Manual Implementation** | **AI-Assisted Implementation** |
|---|---|---|
| Data Source | Human-interpreted PLC documents | Parsed digital PLC documents |
| Conflict Detection | Expert judgement | Automated polarity recognition |
| Time Efficiency | Moderate to low | High |
| Transparency | Full (documented rationale) | Moderate (model-extracted logic) |

| Attribute | Manual Implementation | AI-Assisted Implementation |
|---|---|---|
| Data Requirement | Low (accepts qualitative) | High (requires structured text) |
| Suitability | Early design or low-data projects | Large-scale or data-rich systems |

Both approaches maintain the same scoring and benchmarking procedures, ensuring comparability of results. Together, they allow scalability from single-product studies to multi-system analyses without altering the underlying logic.

## 5. Case Studies

Three case studies were conducted, one each in the agricultural, textile and energy sectors, to evaluate CONFARM's adaptability and robustness. These sectors were chosen to represent variation in system complexity, data structure, and lifecycle scale. Across them, CONFARM generated five, fifteen, and twenty-five sustainability conflicts respectively, reflecting differences in system scope and data detail.

### 5.1 Agriculture: Rice and Corn Systems

This case applies CONFARM to rice and corn cultivation to evaluate sustainability conflicts in agricultural production. These crops were selected because they are widely grown staples with contrasting practices, resource demands and social implications. The aim is to determine which system exhibits more sustainability conflicts across environmental, economic and social pillars, and how these conflicts change with increasing analytical depth. Both manual and AI-assisted implementations (ChatGPT and Gemini) were used to identify decisions, map consequences and score resulting trade-offs. The comparative results show which system has higher conflict counts, stronger cumulative impacts and higher sustainability ratios, demonstrating how CONFARM can support decision-making between agricultural alternatives.

#### 5.1.1 Lifecycle Data and Sources

The lifecycle for both crops was structured into five stages, consistent with established agricultural classifications [14, 15]. Data on energy inputs, resource use and emissions were consolidated from sectoral lifecycle documentation and open agricultural databases [16-18]. The five stages and their representative activities are summarised in Table 4.

Table 4. Product-lifecycle stages and representative activities for corn and rice systems.

| Lifecycle Stage | Description | Representative Activities Considered |
|---|---|---|
| 1. Raw Material Acquisition | Inputs essential for initiating cultivation. | Seed selection, seed treatment, fertiliser and agrochemical sourcing. |
| 2. Manufacturing (Cultivation and Land Preparation) | Core production activities transforming inputs into biomass. | Ploughing, tillage, transplanting, irrigation setup, bund construction. |
| 3. Transportation and Distribution | Handling and movement of materials and resources. | Transport of seeds, fertilisers, and harvested grains. |

| Lifecycle Stage | Description | Representative Activities Considered |
|---|---|---|
| 4. Use Phase (Growth and Maintenance) | Active crop growth involving resource use and maintenance. | Water and nutrient management, pest and weed control, equipment cleaning. |
| 5. End-of-Life Management | Processes following harvest and storage of produce. | Harvesting, drying, packaging, residue management, waste disposal. |

Given the labour-intensive and community-centred nature of farming systems, previous literature identifies the social dimension as dominant, followed by environmental and economic considerations [19, 20]. Accordingly, pillar weights were defined as Social = 0.75, Environmental = 0.50 and Economic = 0.25 to ensure appropriate representation of livelihood and equity factors within the scoring structure.

**5.1.2 Conflict Identification**

Conflicts were identified by tracing relationships between cultivation or process decisions and their resulting sustainability outcomes across the three pillars. Impact magnitudes were assigned using the categorical scale described in Table 5.

Table 5. Impact classification for agricultural sustainability effects.

| Impact Level | Score (i) | Definition | Typical Examples |
|---|---|---|---|
| High | 0.75 | Irreversible or cross-system effects requiring urgent mitigation. | High energy consumption, toxic emissions, major labour exploitation, resource depletion. |
| Medium | 0.50 | Controllable impacts that may escalate cumulatively. | Excess packaging, increased costs, moderate emissions, water overuse. |
| Low | 0.25 | Localised or reversible impacts with minimal systemic risk. | Minor delays, noise, or small-scale inefficiencies. |

Representative conflict sets for rice cultivation were extracted manually and through AI-based processing, as summarised in Tables 6 and 7.

Table 6. Representative five-conflict set for rice cultivation- manual extraction.

| Decision | Positive Effect | Negative Effect |
|---|---|---|
| High-quality seeds | Increased yield potential (Eco 0.75) | High seed cost limits smallholder access (Soc 0.50) |

| Decision | Positive Effect | Negative Effect |
|---|---|---|
| Mechanised ploughing | Enhanced operational efficiency (Eco 0.50) | Increased GHG emissions (Env 0.75) |
| Flood irrigation | Effective weed control (Env 0.50) | Over-exploitation of water resources (Env 0.75) |
| Manual weeding | Reduced herbicide dependency (Env 0.50) | Labour-intensive process (Soc 0.75) |
| Plastic packaging | Protects grains during storage (Eco 0.25) | Plastic waste generation (Env 0.50) |

Table 7. Representative five-conflict set for rice cultivation – AI-based extraction (taken from both ChatGPT and Gemini).

| Decision | Positive Effect | Negative Effect |
|---|---|---|
| High-quality seed use | Higher yield stability (Eco 0.75) | Increased input dependency (Eco 0.50) |
| Mechanised ploughing | Reduced manual labour (Soc 0.50) | Fuel-based emissions (Env 0.75) |
| Controlled irrigation | Improved water efficiency (Env 0.50) | Methane emissions (Env 0.75) |
| Manual weeding | Preserves soil quality (Env 0.50) | Worker fatigue (Soc 0.75) |
| Plastic packaging | Reduced grain losses (Eco 0.25) | Non-biodegradable waste accumulation (Env 0.50) |

Figures 5-6 illustrate the cause-effect (CE) mappings obtained from manual and LLM-based extractions (ChatGPT is taken as an example here).

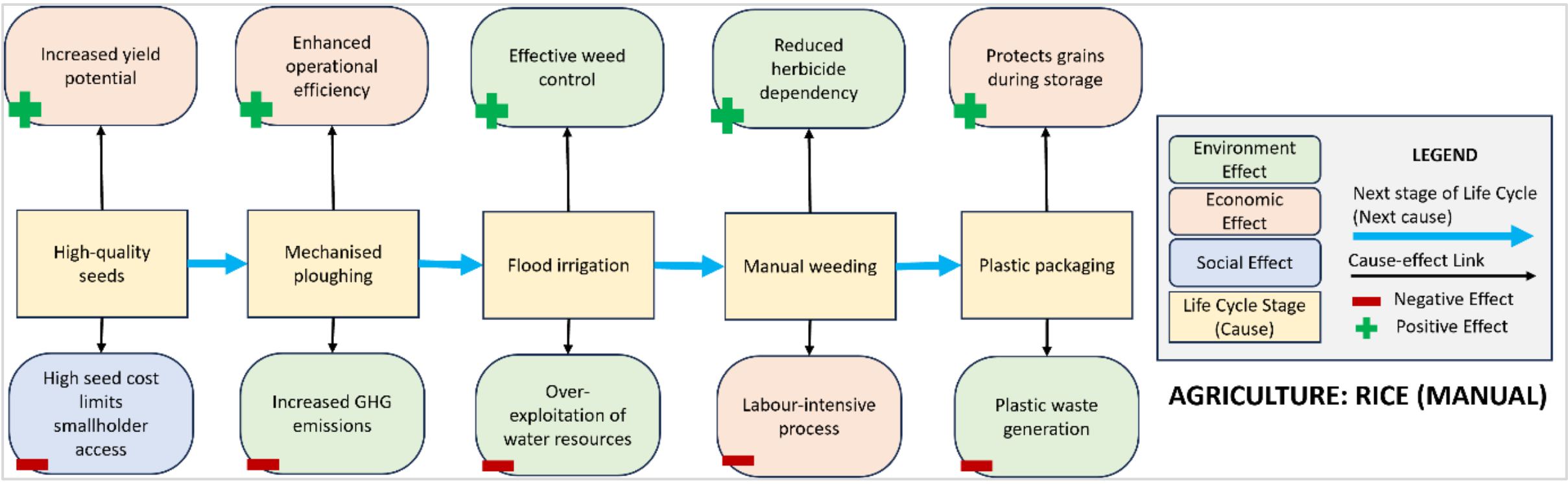

Figure 5. Cause–effect graph for rice cultivation – manual extraction

Figure 5 shows the cause-effect graph constructed through expert review. Rectangular nodes represent cultivation or processing decisions (e.g., irrigation, weeding, packaging), and circular nodes represent the resulting sustainability outcomes. Positive and negative effects are indicated beside each outcome. The graph highlights how a single decision, such as mechanised ploughing or flood irrigation, can produce benefits in one pillar while introducing burdens in another.

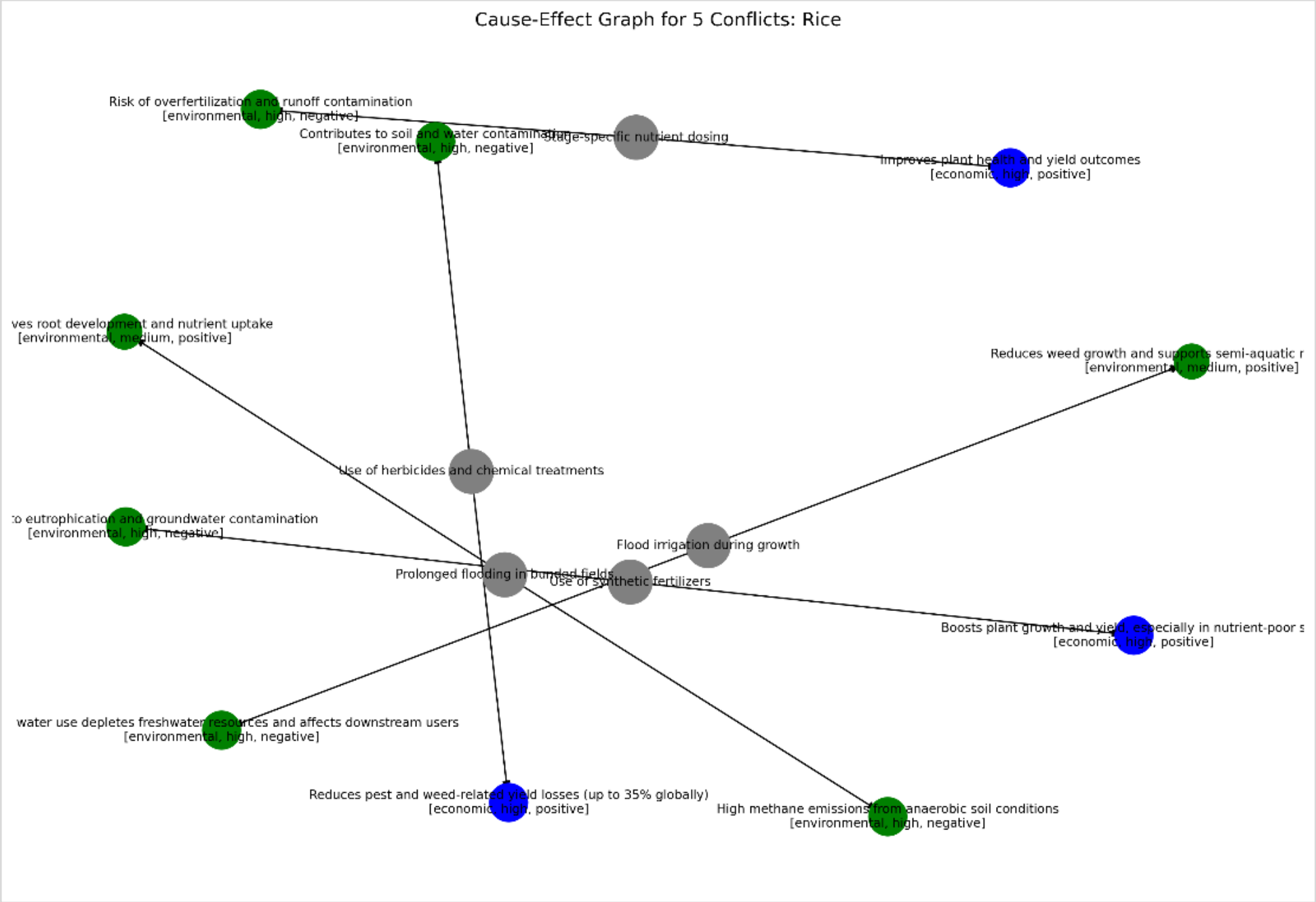

Figure 6. Cause–effect graph for rice cultivation – ChatGPT-based extraction.

Figure 6 presents the same structure generated through AI-assisted parsing of lifecycle documents. Decision and effect nodes are identified automatically, and polarity is inferred from contextual patterns in the text. As in the manual version, each decision is linked to divergent outcomes, showing where sustainability conflicts arise. The similarity to Figure 5 demonstrates that the automated method can replicate the logic of manual cause-effect reasoning, supporting consistent conflict detection across the lifecycle.

### 5.1.3 Scoring Results – 5-Conflict Level

The sustainability scores computed for the five-conflict level is presented in Table 7. The scores depict broadly consistent performance across the three implementations.

Table 8. Sustainability scores for rice cultivation (5-conflict level).

| **Method** | **R = N / T** | **Category** |
|---|---|---|
| Manual | 0.574 | Moderate Sustainability |
| ChatGPT | 0.638 | Moderate Sustainability |
| Gemini | 0.658 | Moderate Sustainability |

These results are interpreted using the magnitude categories and scoring expression introduced earlier (Table 2), which define how weighted effects are aggregated into the sustainability ratio.

#### 5.1.4 Scoring Results – 15-Conflict Level

Extending the analysis to fifteen conflicts introduced additional lifecycle trade-offs related to irrigation efficiency, fertiliser dependency, and energy demand. Results are shown in Table 7.

Table 9. Sustainability scores for rice cultivation (15-conflict level).

| Method | R = N / T | Category |
|---|---|---|
| ChatGPT | 0.579 | Moderate Sustainability |
| Gemini | 0.538 | High Sustainability |

#### 5.1.5 Scoring Results – 25-Conflict Level

At twenty-five conflicts, CONFARM captured more granular interactions, including nutrient run-off, labour-equity implications, and packaging recyclability. The corresponding scores are reported in Table 10.

Table 10. Sustainability scores for rice cultivation (25-conflict level).

| Method | R = N / T | Category |
|---|---|---|
| ChatGPT | 0.565 | High Sustainability |
| Gemini | 0.485 | High Sustainability |

#### 5.1.6 Corn Cultivation

CONFARM was similarly applied to corn cultivation using lifecycle documentation verified through EOS Data Analytics [14] and sectoral agricultural LCA sources [16–18]. Table 11 lists the computed sustainability ratios for increasing conflict densities.

Table 11. Sustainability scores for corn cultivation.

| Conflict Set Size | Manual (R) | Category for Manual | ChatGPT (R) | Category for ChatGPT | Gemini (R) | Category for Gemini |
|---|---|---|---|---|---|---|
| 5 | 0.719 | Low Sustainability | 0.659 | Low Sustainability | 0.703 | Low Sustainability |
| 15 | – | - | 0.646 | Low Sustainability | 0.562 | High Sustainability |
| 25 | – | - | 0.600 | Moderate Sustainability | 0.518 | High Sustainability |

These quantitative outputs provide the empirical basis for cross-case comparison and are further analysed in Section 6.

**5.2 Fashion: Slow vs Fast Fashion**

This case compares slow- and fast-fashion systems. Fast fashion is defined by high-volume, low-cost production and rapid design–manufacturing cycles, whereas slow fashion emphasises durability, smaller batches and reduced throughput. The objective is to examine whether the greater durability, ethical sourcing and lower production intensity of slow fashion reduce sustainability conflicts relative to fast fashion, which offers affordability and scale but increases material and energy use.

CONFARM was applied to both systems using the same analytical structure as in Section 5.1. Lifecycle documentation was sourced from Carbon Trail and supported by peer-reviewed life-cycle studies.

In fashion, environmental impacts dominate (chemical processing, water use, waste and energy), followed by social factors (labour and supply-chain equity) and economic considerations (production cost and retail margin). Pillar weights were set accordingly: Environmental = 0.75, Social = 0.50 and Economic = 0.25. Each system was evaluated at three levels of detail (5, 15 and 25 conflicts). Sustainability ratios for fast fashion and slow fashion are reported in Tables 12 and 13.

Table 12. Sustainability scores for fast-fashion systems.

| **Conflict Set Size** | **Manual (R)** | **Category for Manual** | **ChatGPT (R)** | **Category for ChatGPT** | **Gemini (R)** | **Category for Gemini** |
|---|---|---|---|---|---|---|
| 5 | 0.620 | Moderate Sustainability | 0.750 | Low Sustainability | 0.779 | Low Sustainability |
| 15 | – | - | 0.731 | Low Sustainability | 0.750 | Low Sustainability |
| 25 | – | - | 0.690 | Low Sustainability | 0.741 | |

Table 13. Sustainability scores for slow-fashion systems.

| **Conflict Set Size** | **Manual (R)** | **Category for Manual** | **ChatGPT (R)** | **Category for ChatGPT** | **Gemini (R)** | **Category for Gemini** |
|---|---|---|---|---|---|---|
| 5 | 0.352 | Moderate Sustainability | 0.297 | High Sustainability | 0.449 | Moderate Sustainability |
| 15 | – | - | 0.273 | High Sustainability | 0.444 | Moderate Sustainability |
| 25 | – | - | 0.286 | High Sustainability | 0.424 | Moderate Sustainability |

### 5.3 Energy Systems: Nuclear vs Natural Gas

CONFARM was next applied to the energy sector to compare nuclear and natural-gas electricity generation systems. The same analytical sequence described in Section 5.1 was followed. Lifecycle datasets for both systems were obtained from the Ecoinvent 3.10 database via the EcoQuery interface [29] and structured in accordance with ISO 14040 and ISO 14044 guidelines [27, 28]. These datasets provide inventory data on fuel extraction, power generation, distribution and waste management under the cut-off system model, ensuring consistency with global LCA standards.

Energy systems are characterised by substantial environmental impacts and relatively stable economic and social dimensions. The literature therefore supports assigning the highest weighting to environmental factors, followed by economic and social dimensions [30, 31]. Accordingly, pillar weights were set to Environmental = 0.75, Economic = 0.50 and Social = 0.25. Each system (nuclear and natural gas) was evaluated at three abstraction levels (5, 15 and 25 conflicts). Computed sustainability ratios for nuclear and natural-gas systems are summarised in Table 14 and Table 15, respectively.

Table 14. Sustainability scores for nuclear electricity generation.

| Conflict Set Size | Manual (R) | Category for Manual | ChatGPT (R) | Category for ChatGPT | Gemini (R) | Category for Gemini |
|---|---|---|---|---|---|---|
| 5 | 0.625 | Moderate Sustainability | 0.668 | Low Sustainability | 0.661 | Moderate Sustainability |
| 15 | – | - | 0.640 | Moderate Sustainability | 0.622 | Moderate Sustainability |
| 25 | – | - | 0.601 | Moderate Sustainability | 0.595 | Moderate Sustainability |

Table 15. Sustainability scores for natural-gas electricity generation.

| Conflict Set Size | Manual (R) | Category for Manual | ChatGPT (R) | Category for ChatGPT | Gemini (R) | Category for Gemini |
|---|---|---|---|---|---|---|
| 5 | 0.710 | Low Sustainability | 0.697 | Low Sustainability | 0.703 | Low Sustainability |
| 15 | – | - | 0.660 | Moderate Sustainability | 0.638 | Moderate Sustainability |
| 25 | – | - | 0.618 | Moderate Sustainability | 0.605 | Moderate Sustainability |

The following section presents a comparative discussion and interpretation of the results, highlighting observed trends, sector-specific sensitivities and the consistency between manual and AI-assisted assessments across increasing conflict densities.

## 6. Discussion

This section integrates results from the agricultural, fashion and energy sectors to evaluate framework performance under different data conditions and conflict densities. It also examines alignment with reported sustainability findings and the consistency between manual and AI-assisted assessments in identifying cross-pillar trade-offs.

Across all sectors, comparative patterns were consistent between manual and AI-based implementations. Since only one manual evaluation was conducted per sector, consistency refers not to repeated manual scoring but to the fact that both approaches led to the same relative conclusions about which systems had higher or lower sustainability conflicts. Configurations with higher conflict counts and weighted magnitudes in the manual analysis also produced higher sustainability ratios in both ChatGPT and Gemini outputs. Manual scoring tended to produce slightly lower ratios, reflecting expert interpretation of qualitative and social factors that were less explicit in documentation. AI-based scoring was more uniform where data was well-structured. Consistency was therefore in rank-order agreement rather than numerical equivalence.

Within each sector, results followed expected sustainability hierarchies: slow fashion outperformed fast fashion, nuclear outperformed natural gas on operational emissions, and rice produced a more balanced outcome than corn when social weighting was included. Minor deviations occurred where boundary conditions or weighting distributions varied, but remained within acceptable analytical ranges.

At five conflicts, a few high-magnitude trade-offs dominated and results varied widely. For rice, R ranged from 0.574–0.658 (Table 8), and for fast fashion from 0.620–0.779 (Table 12). Increasing to fifteen conflicts introduced additional interactions and narrowed this spread. Slow-fashion shifted from 0.352–0.449 to 0.273–0.444 (Table 13), while corn converged from 0.659–0.719 to 0.562–0.646 (Table 11). At twenty-five conflicts, all cases stabilised, with differences below 5% for rice (0.565 vs 0.485, Table 10) and slow fashion (0.286 vs 0.424, Table 13). This convergence indicates that CONFARM becomes system-stable once interconnections are sufficiently mapped across lifecycle stages. The progression from five to twenty-five conflicts therefore reflects a transition from surface-level assessment to more holistic evaluation.

### 6.1 Agriculture: Rice and Corn Systems

In agriculture, manual and AI-based assessments showed consistent patterns. With agriculture-specific weights (Social = 0.75, Environmental = 0.50, Economic = 0.25), rice recorded lower R values than corn at the five-conflict level. As conflict density increased to fifteen and twenty-five, ratios declined and converged, stabilising around 0.52–0.58 for rice and ~0.60 for corn. This reflects compensatory lifecycle effects captured through broader mapping.

The ranking, where rice performs slightly better overall, stems from social weighting and management practice. Smallholder rice cultivation contributes to employment and community welfare, while sustainable irrigation and residue management moderate environmental impacts. Corn typically involves greater mechanisation and fertiliser use, improving economic efficiency but reducing social inclusiveness. These patterns align with comparative studies showing rice outperforming maize in integrated social–environmental evaluations when labour equity and resource access are considered [32–33].

Regional assessments support this outcome: in China and India, rice shows higher employment indices and community value-addition despite greater methane emissions [32]. Brazilian and South Asian studies report mechanised maize systems scoring lower on social inclusion and livelihood support [33–34]. Multi-country LCAs find that when labour welfare and resource equity are prioritised, rice ranks equal to or slightly above maize on composite scales [35]. This corroborates CONFARM's weighting approach.

Convergence at fifteen and twenty-five conflicts matches lifecycle literature, where expanding system boundaries stabilises indicator variance and increases comparability [27, 28]. Despite higher methane emissions in environment-focused LCAs, the tri-pillar framework shows that rice can achieve a lower overall R when social and management benefits are included [32, 35].

### 6.2 Fashion Systems: Fast and Slow Production

In the energy sector, CONFARM showed the highest consistency, supported by structured lifecycle data. Nuclear and natural-gas systems were evaluated across extraction, processing, generation, distribution and waste management.

Nuclear energy showed slightly declining R values as conflict density increased (0.625 to ~0.60), while natural gas remained higher (0.71–0.60). This decline reflects cumulative effects captured through conflict mapping, including compensatory efficiencies such as waste-heat recovery. Nuclear systems offered operational advantages in low carbon output, but waste handling and decommissioning lowered performance, consistent with lifecycle studies [40, 41]. Natural gas performed better economically yet faced persistent environmental penalties from methane leakage, combustion emissions and transmission losses [42, 43].

Pillar weights (Environmental = 0.75, Economic = 0.50, Social = 0.25) reflected sector priorities and aligned with ISO 14040/14044 guidance and comparative research on environmental dominance in energy trade-offs [40–44]. Manual assessments were more attentive to end-of-life and regulatory aspects, while AI scoring provided consistent quantification. ChatGPT captured impact variability more finely; Gemini maintained stable scaling.

The results confirm that CONFARM reliably identifies sustainability trade-offs in complex, data-rich domains, while revealing the limitation of static temporal treatment. Incorporating time-dependent parameters would better represent accumulation or decay of long-term environmental impacts [41, 45]. Overall, the energy case demonstrates CONFARM's ability to balance environmental and economic concerns in infrastructure-intensive systems.

### 6.3 Energy Systems: Nuclear and Natural Gas

Pillar weights (Environmental = 0.75, Economic = 0.50, Social = 0.25) reflected sector priorities and aligned with ISO 14040/14044 guidance and comparative research on environmental dominance in energy trade-offs [40–44]. Manual assessments were more attentive to end-of-life and regulatory aspects, while AI scoring provided consistent quantification. ChatGPT captured impact variability more finely; Gemini maintained stable scaling.

The results confirm that CONFARM reliably identifies sustainability trade-offs in complex, data-rich domains, while revealing the limitation of static temporal treatment. Incorporating time-dependent parameters would better represent accumulation or decay of long-term environmental impacts [41, 45]. Overall, the energy case demonstrates CONFARM's ability to balance environmental and economic concerns in infrastructure-intensive systems.

6.4 Comparative Insights Across Sectors

Across agriculture, fashion, and energy, CONFARM demonstrated stable convergence of sustainability ratios with increasing conflict counts. Manual and AI based assessments showed greater deviation in sectors with qualitative documentation, such as agriculture and fashion, while energy, supported by standardised data, exhibited near identical results across methods.

At higher conflict-densities (15–25), all sectors converged within a narrow sustainability ratio range of 0.5–0.6, confirming that expanded cause–effect mapping reduces bias and improves comparability. This cross-sector consistency supports CONFARM's scalability from socially intensive to infrastructure-intensive domains. It also validates the balanced weighting strategy, showing that the tri-pillar approach remains robust across variable data structures and impact contexts.

## 7. Limitations and Future Work

CONFARM developed in this study demonstrates strong adaptability across sectors, although several limitations provide opportunities for refinement.

First, conflicts are currently evaluated as static relationships between decisions and outcomes. Time-dependent effects such as cumulative emissions, resource depletion and delayed social benefits are not represented. Introducing dynamic parameters or system-dynamics modelling would capture long-term variation more realistically.

Second, the AI-based implementation reduces assessment time substantially but depends on the quality and structure of documentation. Incomplete or poorly formatted text can limit conflict extraction. Domain-specific language models and ontology-based sustainability databases would improve both accuracy and contextual interpretation.

Third, the scoring method assumes linear pillar weighting. In practice, priority varies by sector and stakeholder. Adaptive or participatory weighting approaches would allow CONFARM to reflect regulatory requirements, local preferences and evolving societal expectations.

Fourth, although both manual and AI-based routes were effective, social and ethical aspects were often underrepresented, particularly in agriculture and fashion. Stronger integration of participatory data, stakeholder engagement and social-impact metrics would improve inclusiveness and represent community-level dynamics more consistently.

As more cases are added, the conflict database will allow comparison across similar decision contexts and support inference based on precedent rather than subjective judgement. The current work establishes the population method and structure; scalable cross-referencing and benchmarking will emerge as the database grows.

Validation across agriculture, fashion and energy shows that sustainability conflicts can be systematically detected, quantified and compared. CONFARM combines cause–effect reasoning with multi-criteria scoring and AI-based automation, creating a bridge between qualitative insights and quantitative evaluation. Its transparency and adaptability support cleaner-production goals and early-stage design decisions.

Future extensions should incorporate temporal modelling, adaptive weighting and richer social data. Integrating absolute-sustainability metrics such as planetary boundaries and science-based targets will further align results with global objectives. Embedding CONFARM within digital-twin or product-lifecycle management environments would enable proactive conflict detection at scale, supporting decision-making, education and policy evaluation across sectors.

Over time, such integration can shift sustainability assessment from reactive evaluation toward proactive design guidance, enabling products and systems that operate within environmental and social limits.

**8. Competing interests**

The author declare none.

**9. Data availability**

The data that support the findings of this study are openly available at reference numbers [14-26].

**10. Funding Statement**

This work received no specific grant from any funding agency, commercial or not-for-profit sectors.